\documentclass[12pt]{article}
\usepackage{graphics}
\usepackage{amsmath}

\newcommand{\beq}{\begin{equation}}
\newcommand{\eq}{\end{equation}}
\newcommand{\pa}{\partial}

\newcommand{\ub}{_}
\newcommand{\up}{^}
\newcommand{\mb}{\mathbf}
\newcommand{\x}{\times}
\newcommand{\boi}{{\bf \Omega}_i}
\newcommand{\bfvi}{{\bf v}_i}

\begin{document}

\title{\bf Hall Magnetohydrodynamics with Electron Inertia}         
\author{Bhimsen K. Shivamoggi\\
University of Central Florida}        
\date{\today}          
\maketitle

\noindent
{\bf Abstract}

Hall magnetohydrodynamics with electron inertia is considered. A much wider class of equilibrium solutions and the concomitant self-organization aspects are discussed. The force-free field state {\textbf B} $\sim$ {\textbf J} is shown to be a sufficient condition to satisfy this generalized class of equilibria.

\section{Introduction}       
Hall magnetohydrodynamics (Drake \cite{Dra}) is characterized by two disparate length scales: the macroscopic scale $\ell$ of the system and the ion skin depth $d\ub{i}\equiv c/\omega\ub{pi}$, with $\ell < d_i.$ On the other hand, in cases where $\ell < d\ub{e},$ electron inertia effects become important. This regime is the subject of discussion here pertaining to a much wider class of equilibrium solutions and the concomitant self-organization aspects in the dynamics.

\section{Governing Equations}
The equation of motion of the ions (in usual notation) is 

\beq
\frac{\pa \mb{V}_i}{\pa t} - \frac{e}{m_i}\mb{E} - \mb{V}_i \x \left(\boi + \frac{c}{m_i c} \mb{B}\right) = -\nabla\left( \frac{\mb V\up{2}}{2}+P\ub{i}\right)
\eq

\noindent
which may be rewritten as 

\beq
\frac{\pa}{\pa t} \left(\mb{V}_i+ \frac{e}{m\ub{i}c} \mb{A}\right)- \mb{V}\ub{i}\x\left(\boi+\frac{e}{m\ub{i}c} \mb{B}\right)=-\nabla\left(\frac{\mb{V}\ub{i}\up{2}}{2}+P_i\right)
\eq

\noindent
where,

\beq
\mb{E} \equiv - \frac{1}{c}\frac{\pa\mb{A}}{\pa t},\ \boi\equiv\nabla\x\bfvi,
\eq

\noindent
and we have assumed barotropic conditions - 

\beq
\nabla P_{e,i}\equiv\frac{1}{n_{e,i}}\nabla p_{e,i}.
\eq

Upon taking the curl of equation (2), we obtain
 
\beq
\frac{\pa}{\pa t}\left( \mb{\Omega}_i + \frac{e}{m\ub{i}c}\mb{B}\right) - \nabla \x \left[ \mb{V}\ub{i}\x\left(\mathbf{\Omega}\ub{i}+\frac{e}{m\ub{i} c}\mb{B}\right)\right]=\mb{0}
\eq

The equation of motion of the electrons is 
\beq
\frac{\pa\mb{V}\ub{e}}{\pa t} + \frac{e}{m\ub{e}} \mb{E} - \mb{V}\ub{e} \x \left(\nabla \x \mb{V}\ub{e}- \frac{e}{m_e c} \mb{B}\right) = - \nabla \left(\frac{\mb{V}_e\up{2}}{2}+ P\ub{e}\right).
\eq

Assuming the quasi-neutrality situation

\beq
n_e\approx n_i=n
\eq

\noindent
and noting that the total current density is given by 

\beq
\mb{J}=ne(\mb{V}\ub{i}-\mb{V}\ub{e})
\eq 

\noindent
equation (6) may be rewritten as

\beq
\frac{\pa}{\pa t} \left[\left(\mb{V}\ub{i}- \frac{\mb J}{ne}\right) - \frac{e}{m\ub{e}c} \mb{A}\right] - \left( \mb{V}\ub{i}-\frac{\mb J}{ne}\right) \x\left[\nabla \x \left(\mb{V}\ub{i}-\frac{\mb{J}}{ne}\right)-\frac{e}{m\ub{e}c}\mb{B}\right]=-\nabla\left(\frac{\mb{V}\ub{e}\up{2}}{2}+P\ub{e}\right)
\eq

Upon taking the curl of equation (9), we obtain

\beq
\frac{\pa}{\pa t}\left(\mb{\Omega}\ub{i}- \frac{e}{m_e c}\mb{B}_e\right)-\nabla \x\left[\left( \mb{V}_i - \frac{\mb{J}}{ne}\right) \x\left(\mb{\Omega}\ub{i}-\frac{e}{m_e c}\mb{B}_e\right)\right]=\mb{0}
\eq

\noindent
where,

\beq
\mb{B}_e \equiv \mb{B}- d_e^2\nabla\up{2} \mb{B}, \ \ d_e^2\equiv\frac{c\up{2}}{\omega_{pe}^2}.
\eq

\section{Invariants}
Equations (2) and (8) admit the following invariants:
\medskip

\begin{align}
Total\  energy: E               & = \int_V\left[ \mb{V}_i^2+\mb{B}^2+\left(\mb{V}_i - \frac{\mb{J}}{ne}\right)^2\right]d \mb{x}\\
Ion\  helicity:H_i      & = \int_V\left(\mb{V}_i + \frac{e}{m_i c}\mb{A}\right)\cdot \left(\mb{\Omega}_i + \frac{e}{m_i c}\mb{B}\right) d\mb{x}\\
Electron\  helicity:H_e & = \int_V \left(\mb{V}_i - \frac{e}{m_e c}\mb{A}_e \right)\cdot\left(\mb{\Omega}_i - \frac{e}{m_e c}\mb{B}_e\right) d\mb{x}
\end{align}

\noindent
where $V$ is the volume occupied by the plasma and 

\beq
\mb{A}_e \equiv \mb{A}- d_e^2\nabla^2\mb{A}.
\eq

\section{Beltrami Conditions}

The plasma Beltrami conditions that correspond to stationary solutions of equations (5) and (10) are 

\beq
\mb{\Omega}_i + \frac{e}{m_i c} \mb{B} = a\mb{V}_i
\eq

\beq
\mb{\Omega}_i - \frac{e}{m_e c}\mb{B}_e = b\left(\mb{V}_i - \frac{\mb{J}}{ne}\right)
\eq

\noindent
where a and b are constants. Equations (16) and (17) may also be obtained by minimizing $E$ while keeping $H_i$ fixed and $H_e$ fixed, respectively.

Combining equations (16) and (17), we obtain

\beq
\frac{1}{ne}\nabla \x (\nabla \x \mb{J})- \frac{1}{ne}(a\  +\  b)(\nabla \x \mb{J}) + \left[\left(\frac{e}{m_i c^2} + \frac{e}{m_e c^2}\right) + \frac{ab}{ne} \right] \mb{J} - \frac{e}{m_e c} \left( a + b\frac{m_e}{m_i}\right)\mb{B} = \mb{0}
\eq

\noindent
or 

\beq
d_e^2 \nabla \x (\nabla \x \mb{J}) - d_e^2 (a\ +\ b)(\nabla \x \mb{J}) + \left(1+ \frac{m_e}{m_i}+ abd_e^2\right)\mb{J} - \left(a + b\frac{m_e}{m_i}\right)c\mb{B}= \mb{0}.
\eq

Equation (19) is, of course, identically satisfied by the {\it force-free} field $ \mb{B} \sim \mb{J}$ (Woltjer \cite{Wol})! If on the other hand, electron inertia is neglected, equation (19) leads, as to be expected, to the force-free field condition -

\beq
\mb{J} - ac\mb{B}= \mb{0}
\eq

\noindent
or

\beq
\nabla \x \mb{B} - a\mb{B} = \mb{0}.
\eq

Using (16), equation (21) becomes 

\beq
\frac{1}{a^2} \nabla \x (\nabla \x \mb{V}_i)  - \frac{2}{a} (\nabla \x \mb{V}_i) + \mb{V}_i = \mb{0}.
\eq

Putting,

\beq
\varepsilon \equiv -\frac{1}{a}
\eq

\noindent
equation (22) becomes

\beq
\varepsilon^2 \nabla \x (\nabla \x \mb{V}_i ) + 2\varepsilon (\nabla \x \mb{V}_i) + \mb{V}_i= \mb{0}
\eq

\noindent
which is the same as the one given by Mahajan and Yoshida \cite{Mah}, on specifying their constants $a$ and $b$ as follows:

\beq
a=\frac{1}{2} \ \ \ \ \ and \ \ \ \ b=0.
\eq

On the other hand, in the electron-inertia dominated regime, equation (19) yields

\beq
\nabla \x {\bf J} - (a + b) {\bf J} + ab{\bf B} = {\bf 0}
\eq

which is the same as the one pertaining to the electron MHD (EMHD) regime given previously (Shivamoggi \cite{Shi}).

\section{Discussion}

In view of the emergence of a significant class of exact solutions of the plasma dynamics equations under the Beltrami condition and their correlation to real plasma behavior, one wonders whether plasmas have an intrinsic tendency towards Beltramization. Though we do not have good understanding of this aspect, it is known that Beltramization provides the means via which the underlying system can accomplish 

\bigskip
\ \ \ \ *ergodicity of the streamlines of the respective flows, (Moffatt \cite{Mof})

\bigskip
\ \ \ \ *selective dissipation of kinetic energy (Woltjer \cite{Wol}).

On the other hand, in the Beltrami states, given by (16) and (17), we obtain from equations (2) and (9) the {\it Bernoulli} conditions - 

\begin{subequations}
\beq
P_i + \frac{1}{2}{\bf v}_i^2 = const,\ \forall{\bf x}\in V
\eq

\beq
P_e + \frac{1}{2}{\bf v}_e^2 = const, \ \forall{\bf x}\in V
\eq
\end{subequations}

\noindent
as in the hydrodynamic case. This appears to signify manifestation of some common features in the Beltrami states, the diversity of the physics underlying the various fluid and plasma models notwithstanding, as previously mentioned in \cite{Shi}.

\end{document}